\documentclass[prd,aps,preprint,groupedaddress,superscriptaddress,amssymb,nofootinbib]{revtex4-2}
\usepackage[british]{babel}
\usepackage[normalem]{ulem}
   
\usepackage{mathrsfs,bm,amsthm}
\usepackage{amsmath}
\usepackage{natbib,textcase}
\usepackage{graphicx,color,soul,comment}
\usepackage[pdftex,pdfusetitle]{hyperref}

\usepackage[capitalize]{cleveref}
\crefname{subequation}{Eqs.}{Eqs.}
\labelcrefformat{subequation}{#2(#1)#3}
\usepackage{subfig}
\usepackage{physics}
\usepackage{array,booktabs} 
\usepackage{multirow} 
\newcolumntype{L}{>{$}p{20mm}<{$}} 
\makeatletter\g@addto@macro\bfseries{\boldmath}\makeatother%

\allowdisplaybreaks

\def\be#1\ee{\begin{align}#1\end{align}}

\renewcommand{\ge}{\geqslant}
\renewcommand{\geq}{\geqslant}

\renewcommand{\leq}{\leqslant}

\usepackage[usenames,dvipsnames]{xcolor}
\hypersetup{colorlinks=true, citecolor=ForestGreen, linkcolor=ForestGreen,urlcolor=ForestGreen}

\begin{document}
\title{Geodesically complete universes}
\author{Ra\'ul Carballo-Rubio}
\email{raul@sdu.dk}
\affiliation{CP3-Origins, University of Southern Denmark, Campusvej 55, DK-5230 Odense M, Denmark}
\author{Stefano Liberati}
\email{liberati@sissa.it}
\affiliation{SISSA - International School for Advanced Studies, Via Bonomea 265, 34136 Trieste, Italy}
\affiliation{IFPU, Trieste - Institute for Fundamental Physics of the Universe, Via Beirut 2, 34014 Trieste, Italy}
\affiliation{INFN Sezione di Trieste, Via Valerio 2, 34127 Trieste, Italy}
\author{Vania Vellucci}
\email{vvellucc@sissa.it}
\affiliation{SISSA - International School for Advanced Studies, Via Bonomea 265, 34136 Trieste, Italy}
\affiliation{IFPU, Trieste - Institute for Fundamental Physics of the Universe, Via Beirut 2, 34014 Trieste, Italy}
\affiliation{INFN Sezione di Trieste, Via Valerio 2, 34127 Trieste, Italy}

\begin{abstract}
Singularity theorems demonstrate the inevitable breakdown of the concept of continuous, classical spacetime under highly general conditions. Quantum gravity is expected to intervene to avoid singularities and models so far hint towards several regularized geometries, in which limited spacetime regions requiring full quantum gravitational description can be safely covered by an extension of some suitable spacetime geometry.  Motivated by these premises, in recent years, a systematic, quantum gravity agnostic, study has been carried out to catalogue all the  conceivable non-singular, continuous, and globally hyperbolic geometries arising from evading Penrose's focusing theorem in gravitational collapse.
In this study, we extend this inquiry by systematically examining all potential non-singular, continuous, and globally hyperbolic extensions into the past of Friedmann–Lemaître–Robertson–Walker (FLRW) metrics. As in the black hole case, our investigation  reveals a remarkably limited set of alternative scenarios. The stringent requisites of homogeneity and isotropy drastically restrict the viable singularity-free geometries to merely three discernible non-singular cosmological spacetimes: a bouncing universe (where the scale factor reaches a minimum in the past before re-expanding), an emergent universe (where the scale factor reaches and maintains a constant value in the past), and an asymptotically emergent universe (where the scale factor diminishes continually, asymptotically approaching a constant value in the past). We also discuss the implications of these findings for the initial conditions of our universe, and the arrow of time.
\end{abstract}

\maketitle
\newpage
\tableofcontents

\section{Introduction}

Penrose's theorem, published in 1965~\cite{Penrose:1964wq}, had an immediate impact in the general relativity (GR) community~\cite{Senovilla:2014gza}. It showed that singularities in gravitational collapse are unavoidable under reasonable physical assumptions.

Shortly after, Hawking pointed out that similar results would hold for cosmological situations~\cite{Hawking:1965mf,Hawking:1966vg,Hawking:1970zqf} (see also~\cite{Geroch:1966ur}), thus showing that the same robustness holds for the singularities characteristic of Friedmann–Lemaître–Robertson–Walker (FLRW) spacetimes. In particular, any universe that is well approximated at large scales by the FLRW spacetime (such as ours~\cite{Green:2014aga}), would be expected to become singular if GR, as well as the assumptions that enter the theorem, were to hold indefinitely backwards in time.

However, the current consensus is that some of these assumptions, and most likely the very framework provided by GR, fails to describe adequately the very early universe, at least because quantum gravitational effects are expected to become relevant before reaching a singularity. Understanding the precise form in which this happens is subject of intense research, with the development of diverse theoretical models~\cite{Ellis:2002we,Mukherji:2002ft,Bojowald:2005epg,Alesci:2016xqa,Brandenberger:2023ver, Chakraborty:2023lav, Battista:2023glw, Geshnizjani:2023hyd, Franco:2024pmn,Lesnefsky:2022fen,Easson:2024uxe,Easson:2024fzn,Barca:2021wmj,Barca:2023epu} that could eventually be tested observationally~\cite{Novello:2008ra,Brandenberger:2017pjz,Gozzini:2019nbo,Garay:2013dya}. 

Here, we extend a, quantum gravity agnostic,   geometric analysis already successfully carried out for classifying the possible geometries of regular black holes~\cite{Carballo-Rubio:2019nel,Carballo-Rubio:2019fnb}, to classify all the non-singular, continuous, and globally hyperbolic extensions into the past of  FLRW metrics. In doing so, we shall find that a limited catalogue of alternative scenarios can be realized. Still such scenarios are sufficiently general to encompass all of the models previously illustrated in the quantum gravity literature, and illuminate on the possible initial conditions for our universe and their implication for the time arrow problem.

The paper is organized as follows: in section~\ref{sec:cases} we discuss the assumptions at the base of cosmological singularity theorems. In section~\ref{sec:TrapFRW} we discuss the structure of trapped regions in FLRW geometries and its relevance for our analysis. In section~\ref{sec:flat} we then analyze, for flat or open FLRW universes, the possible alternatives to a past spacelike singularity such as the ``Big Bang". In section~\ref{sec:Closed} we extend the same analysis to the case of closed FLRW universes. We close with a discussion of the implications of our findings in section~\ref{sec:Concl}.

\section{Beyond Penrose's theorem \label{sec:cases}}
\def\defocus{{\mathrm{defocus}}}

Penrose's 1965 theorem~\cite{Penrose:1964wq} marked a milestone in our understanding of gravity~\cite{Senovilla:2014gza}. It was based on a quite restricted set of assumptions that can be summarized as
\begin{itemize}
\item The ``null convergence condition", $R_{ab}k^a k^b \ge 0\: \:\forall$ null vector $k^a$, holds (this implies in GR the null energy condition).
\item The manifold is a time orientable, globally hyperbolic spacetime, ${\cal M }= R \times \Sigma^3$ admitting a non-compact Cauchy hypersurface.
\item At some point in a gravitational collapse process, a closed future-trapped surface ${\mathcal T}^2$ forms, with a (negative) maximum expansion $\theta_\mathrm{max} = \theta_0 < 0$.
\end{itemize}
The theorem then proceeds in two steps. The first one consists in showing that an initial negative expansion and the null convergence condition are enough to prove, via the Raychaudhuri equation, that the expansion will become infinitely negative in the future for some finite value of the affine parameter of a null congruence orthogonal to ${\mathcal T}^2$. This focusing point is expected to be singular. The second step consisted in a ingenious use of topological arguments to prove that assuming the regularity of the focusing point 
is incompatible with the presence of a non-compact Cauchy hypersurface.
 
The original singularity theorem above summarized, actually proves that at the end of a gravitational collapse, either the spacetime is geodesically incomplete or it must develops a Cauchy horizon so that the presence of a non-compact Cauchy hypersurface cannot hold.
This motivated Penrose and Hawking to formulate a second theorem in 1970 not relying on Cauchy hypersurfaces, and hence also extendible to closed universes, where non-compact Cauchy hypersurfaces are absent~\cite{Hawking:1970zqf}. In order to analyze general closed universes, we would need to use this second theorem. However, we will see that, for spatially homogeneous and isotropic spacetimes, the framework provided by the original Penrose theorem suffices to classify the possible non-singular spacetimes.

In this work we analyze and classify globally hyperbolic and time-orientable manifolds describing expanding, homogeneous and isotropic regular universes. In order to avoid the formation of a singularity, at least one of the assumptions of the Penrose theorem must necessarily be violated. For flat and open manifolds the last two assumptions of the theorem will always be verified and thus the geodesically complete geometries that we will find will necessarily violate the null convergence condition. For closed manifolds similar considerations imply that all the geodesically complete geometries that we find necessarily violate the timelike convergence condition (see section \ref{sec:Closed}).

\section{Trapped regions in FLRW geometries}
\label{sec:TrapFRW}

Let us consider the Friedmann-Lemaitre-Robertson-Walker (FLRW) metric~\cite{Weinberg:1972kfs}:
\begin{equation}
 d s^2=-d t^2+a^2(t)\left[d r^2+R^2(r)\left(d \theta^2+\sin ^2 \theta d \phi^2\right)\right],   
 \label{eq:FRWgen}
\end{equation}
with $R(r)=(\sin r, r, \sinh r) \text { if }(k=1,0,-1) $ respectively.

The family of past-directed radial null geodesics in this space-time has tangent
vector field \cite{Ellis:2003mb}:
\begin{equation}
  V_\pm^a=\frac{1}{a(t)}\left(-1, \pm \frac{1}{a(t)}, 0,0\right)=\frac{dx^a}{d\lambda},
  \label{geo}
\end{equation}
where $d\lambda=-a(t) dt \pm a(t)^2 dr$ is an affine parameter\footnote{Note that these two null vectors are defined only up to a multiplicative function of the respectively orthogonal null coordinate. The redefinition $V_{\pm} \rightarrow g(u_{\mp}) V_{\pm}$, with $g(u_{\mp})>0$ and $du_{\pm}=-dt/a(t)\pm dr$, stills results into affinely parameterized vector fields. However, $u_{\pm}$ are constant along the respectively orthogonal geodesics, which implies $V_{\pm}^a\nabla_a g(u_\mp)=0$. Hence, this redefinition changes expansions by a positive multiplicative factor and therefore does not affect our identification of trapped regions.}. Note that the $+$ sign indicates outgoing geodesics, while the $-$ indicates ingoing geodesics. Along such null geodesics, from Eq.~\eqref{eq:FRWgen} we have $dr/dt=\mp 1/a$ and thus $\lambda= - 2 \int a(t) dt$.

The expression for the expansion of such geodesics is then\footnote{The expression provided for the expansion is not the usual one derived as $\partial_t \ln{A}$, where $A$ is the proper cross-sectional area~\cite{Poisson:2009pwt}, but it is that divided by $1/a(t)$ in order to coincide with the divergence of the tangent vector normalized as in Eq.~\eqref{geo}.}
\begin{equation}
 \theta_{\pm} =\frac{2}{a^2(t)} \left[ -\dot{a}(t) \pm \frac{\partial R(r) / \partial r}{R(r)}\right], 
 \label{expansion}
\end{equation}
in terms of the time $t$, or
\begin{equation}
 \theta_{\pm} =\frac{2}{a^2(\lambda)} \left[ 2a(\lambda)a'(\lambda) \pm \frac{\partial R(r) / \partial r}{R(r)}\right], 
 \label{expansionLambda}
\end{equation}
in terms of the affine parameter $\lambda$. In these expressions as well as the rest of the paper, $\dot{a}=da/dt$ and $a'=da/d\lambda$, so that $\dot{a}=-2aa'=-(a^2)'$. Due to the one-to-one relation between $\lambda$ and $t$, we can choose any of these variables to parametrize null geodesics. In the following, we will always use $t$ for this purpose. When working with specific geodesic congruences we will always choose, without loss of generality, a reference point $\lambda=\lambda_\star$ such that $t(\lambda_\star)=0$.

Let us now discuss the structure of trapped surfaces for expanding universes. In the flat and open case, taking into account that $R(r)$ is always a positive and monotonically increasing function in its range of definition, any expanding universe [$\Dot{a}(t)>0$ or  $a'(\lambda)<0$] satisfies
\begin{equation}
    \theta_- \leq 0,
\end{equation}
while 
\begin{equation}
    \theta_+
    \begin{cases}
    >0 &\text{if } \partial \ln R(r)/\partial r>\Dot{a},\\
    \leq 0 &\text{if } \partial \ln R(r)/\partial r\leq \Dot{a}.
    \end{cases}
\end{equation}

In the closed case, $R(r)$ is no more a monotonically increasing function, and can assume negative values. However, we {also} have to take into account that, in such universe, $\theta_+<0$ doesn't correspond necessarily to a trapped region since for some values of $r$ outgoing and ingoing geodesics exchange roles, as we shall elaborate in Sec.~\ref{sec:Closed}. Thus, one has to check the sign of both $\theta_+$ and $\theta_-$, and a trapped region is present only if both of them are negative. We get
\begin{equation}
    \theta_+
    \begin{cases}
    >0 &\text{if } \, \partial \ln R(r)/\partial r>\Dot{a},\\
    \leq 0 &\text{if } \, \partial \ln R(r)/\partial r\leq \Dot{a}.
    \end{cases}
\end{equation}
\begin{equation}
    \theta_-
    \begin{cases}
    >0 &\text{if } \, \partial \ln R(r)/\partial r<-\Dot{a},\\
    \leq 0 &\text{if } \, \partial \ln R(r)/\partial r\geq -\Dot{a}.
    \end{cases}
\end{equation}

\begin{figure}[htb]
    \centering    
\includegraphics[width=0.3\columnwidth]{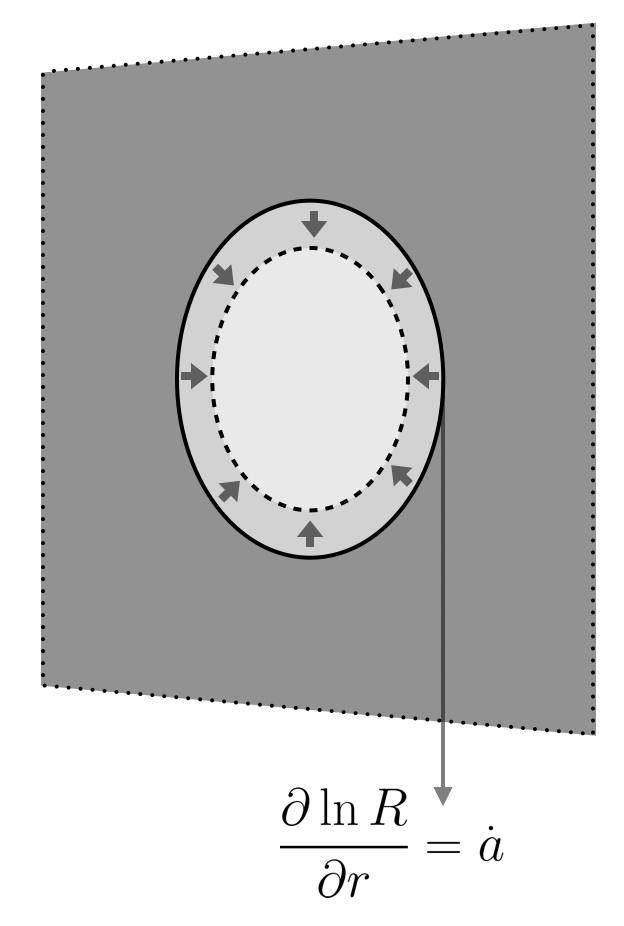}
    \caption{\small Structure of trapped surfaces (dark grey) in a given slice of constant time $t$ of FLRW spacetimes (for open universes, it is necessary that $\dot{a}>1$ for trapped surfaces to exist). The solid circle indicates the location of marginal trapped surface for a given value of $\dot{a}$, and the arrows pointing to the dashed line indicate how this location changes as $\dot{a}$ increases.}
    \label{fig:trsr}
\end{figure}

The function $\partial \ln R(r) / \partial r$ goes from $+ \infty$ at $r=0$ to $0$ at $r=\infty$ for the flat case ($k=0$); from $+\infty$ to $1$ in the open case ($k=-1$); and from $+\infty$ and $-\infty$ for the closed case ($k=1$). Hence, both flat and closed expanding universes always have trapped surfaces with the structure depicted in Fig.~\ref{fig:trsr}, which is also shared by open universes with $\Dot{a}>1$.\footnote{{Note that, as we have chosen $k$ to be dimensionless following standard conventions, $a(t)$ has dimensions of length. As we are also choosing units in which $c=1$, $\dot{a}(t)$ is dimensionless.}}

\begin{figure}[htb]
    \centering
\includegraphics[width=6.5cm]{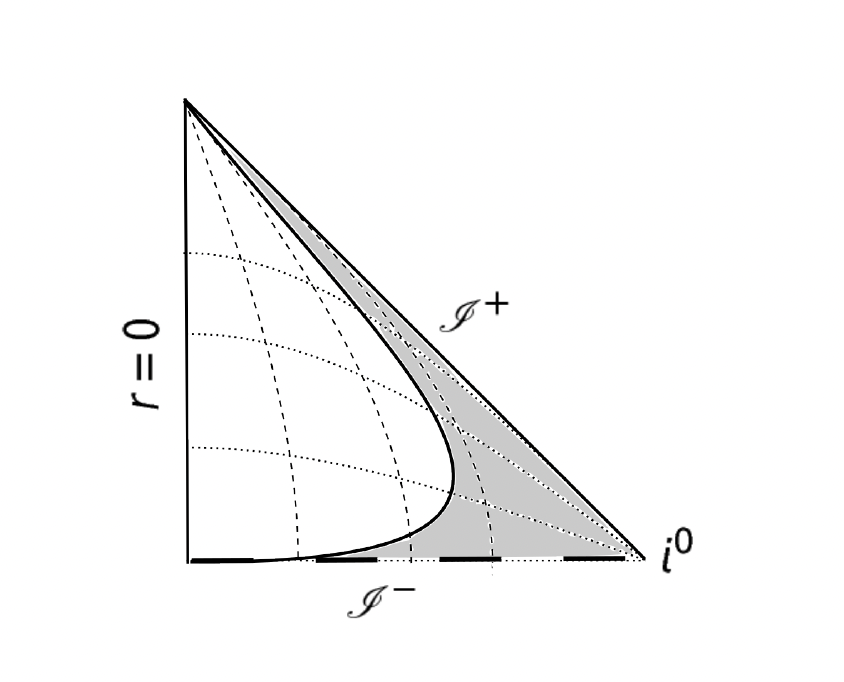}
    \caption{\small Penrose Diagram of a singular decelerating universe. The shaded region is the one containing  trapped surfaces (which, for open universes, requires $\dot{a}>1$ in order to exist). Note that the shading indicates trapped points with respect to a fiduciary reference point which has no physical meaning: due to homogeneity and isotropy, either all points are trapped, or no points are trapped.}
    \label{fig:TrpSurPenrose}
\end{figure}

\cref{fig:TrpSurPenrose} shows the causal structure of a decelerating FLRW metric and its trapped region, which will exist for the open case only for $\Dot{a}>1$. Trapped surfaces appear for sufficiently large distances to the reference point $r=0$ being used. However, due to homogeneity and isotropy, the choice of reference point is fiduciary and has no physical meaning. This is different with respect to spherically-symmetric black holes, in which the gravitational potential has a defined center. Hence, in a FLRW spacetime it is enough to show that a point belongs to a trapped surface to conclude that all points belong to trapped surfaces. In other words, there are no trapped regions in the usual sense: either all points are trapped, or no points are trapped.

Finally, regardless of the spatial curvature and the structure of the trapping surfaces, imposing regularity of curvature invariants implies some further constraints on the nature of the scale factor as a function of time. The Ricci scalar $\mathcal{R}$ takes the form
\begin{equation}\label{eq:Ricciscalar}
\mathcal{R}=\frac{6 \left[k+ a(t) \Ddot{a}(t)+\dot{a}^2(t)\right]}{a(t)^2} ,
\end{equation}
while the Kretschmann scalar $\mathcal{K}$ reads
\begin{equation}
\mathcal{K}=\frac{12\left[a^2(\ddot{a})^2+(k+\dot{a}^2)^2\right]}{a^4}. 
\end{equation}
As a consequence, a necessary condition to avoid curvature singularities is for $a(t)$ to be at least a $C^2$ function.

\section{Null expansions-based classification of regular cosmological spacetimes
}\label{taxonomy}
We shall now analyze the possible conditions under which a singularity in the past of a FLRW universe can be avoided, starting with some general considerations. 

{For flat and open universes (that is, well approximated by either $k=0$ or $k=-1$ FLRW spacetimes), the discussion is paralell to the one in~\cite{Carballo-Rubio:2019fnb}, with the only difference that we will consider congruences of past-directed (instead of future-directed) null geodesics. Hence, the different geometric possibilities for these cases are in one-to-one correspondence to the possible ways in which spacetimes can be deformed to avoid focusing points. For general closed universes, Penrose's theorem does not apply in general, and therefore the classification below would not apply without further considerations. However, as discussed in Sec.~\ref{sec:Closed}, this classification is still meaningful for homogeneous and isotropic closed universes.

As explained in Sec.~\ref{sec:TrapFRW}, moving to bigger values of the affine parameter for these congruences corresponds to move towards the past (i.e., to smaller values of $t$), and in the following equations we will always choose, without loss of generality, a reference point $\lambda=\lambda_\star$ such that $t(\lambda_\star)=0$. As we start with negative expansion for some of the past directed outgoing null geodesics (we start in a past trapped region), the avoidance of a focusing point requires at some point a change of sign for $\theta_+$ and hence that the latter vanishes.

However, differently from the black-hole cases studied in ~\cite{Carballo-Rubio:2019fnb}, for cosmological spacetimes, the expansion does not vanish at the same value of the affine parameter for all null geodesics. We shall name the point at which $\theta_{+}=0$ for the most trapped null geodesic, and thus $\theta_+\geq0$ for all null geodesics, the defocusing point.

Starting from a reference value of the affine parameter $\lambda_*$, we can classify different geometries depending on the behavior of radial null congruences for $\lambda=\lambda_0>\lambda_*$. The rate of change of the area element orthogonal to the congruence is determined by the following equation:
\begin{equation}
\ln\left(\frac{\left.\delta A_+\right|_{\lambda=\lambda_0}}{\left.\delta A_+\right|_{\lambda=\lambda_*}}\right)=\int^{\lambda_0}_{\lambda_*}\text{d}\lambda\,\theta_+(\lambda).
\label{First}
\end{equation}
Note that we are choosing a reference point $ \lambda_* $ for which  $ \delta A_+|_{\lambda=\lambda_*} \neq 0 $.
As it was shown in~\cite{Carballo-Rubio:2019fnb}, a corollary of this equation is that a congruence has a focusing point at a finite affine distance $\lambda=\lambda_0$ if and only if $\left.\theta_+\right|_{\lambda=\lambda_0}=-\infty$. 

In order to avoid that the spacetime is geodesically incomplete we need to modify the spacetime geometry in the vicinity of the focusing point, either creating a defocusing point ($\theta_+=0$) or displacing the focusing point to infinite affine distance.

We can now proceed to a systematic consideration of all the possible regularization of FLRW universes. In doing so we shall use the same categorization used in~\cite{Carballo-Rubio:2019fnb} based on three parameters, i.e.,~the value of the affine parameter $\lambda$ for which we have a defocusing of the outgoing congruence, the value of the radius $R(\lambda)$ of the area element orthogonal to the congruence at that point, and the value of the ingoing congruence expansion at that point, $\bar{\theta}$.

\begin{description}
\item[Case A]{Defocusing point at a finite affine distance, $\lambda_\defocus=\lambda_0$.}
\begin{itemize}
\item[A.I:]{$(\lambda_0,R_0,\bar{\theta}<0)$: 
The expansion $\theta_+$ vanishes and changes sign at a finite affine distance $\lambda=\lambda_0$ or, in terms of the radius of the area element orthogonal to the null outgoing congruence, at a value $R_0={R(\lambda_0)}>0$. On the other hand, the expansion of the intersecting ingoing radial null geodesics remains negative until (and including) $\lambda_0$, so that $\bar{\theta}=\left.\theta_-\right|_{\lambda=\lambda_0}<0$.}

\item[ A.II:]{$(\lambda_0,R_0,\bar{\theta}\geq0)$: 
The only difference with respect to the previous case is that the expansion of the intersecting ingoing radial null geodesics does not remain negative at $R_0={R(\lambda_0)}$, $\bar{\theta}=\left.\theta_-\right|_{\lambda=\lambda_0}\geq0$.}
\end{itemize}

\item[Case B]{Defocusing point at an infinite affine distance, $\lambda_\defocus=\infty$.}
\begin{itemize}
\item[B.I:]{$(\infty,R_{\infty},\bar{\theta}<0)$: 
The expansion $\theta_+$ vanishes in the limit $\lambda\rightarrow\infty$, in a manner such that the integral in Eq. \eqref{First} is convergent. The corresponding asymptotic value of the radius of the area element orthogonal to the outgoing null congruence is $R_{\infty}={\lim_{\lambda_0\rightarrow\infty}R(\lambda_0)}>0$. The expansion of the intersecting ingoing radial null geodesics remains negative there, so that $\bar{\theta}=\left.\theta_-\right|_{\lambda \rightarrow \infty }<0$.}

\item[B.II:]{$(\infty,R_{\infty},\bar{\theta}\geq0)$: 
The only difference with respect to the previous case is that the expansion of the intersecting ingoing radial null geodesics does not remain negative at $R_{\infty}={\lim_{\lambda_0\rightarrow\infty}R(\lambda_0)}$, $\bar{\theta}=\left.\theta_-\right|_{\lambda \rightarrow \infty}\geq0$.}

\item[B.III:]{$(\infty,0,\bar{\theta}<0)$: 
The expansion $\theta_+$ vanishes in the limit $\lambda\rightarrow\infty$, in a manner such that the integral in Eq. \eqref{First} is divergent. Thus, the radius of the area element orthogonal to the congruence vanishes asymptotically along these geodesics (in other words, there is an asymptotic focusing point), $R_\infty=\lim_{\lambda_0\rightarrow\infty}R(\lambda_0)=0$. The expansion of the intersecting ingoing radial null geodesics remains negative at $0$, so that $\bar{\theta}=\left.\theta_-\right|_{\lambda \rightarrow \infty}<0$.}

\item[B.IV:]{$(\infty,0,\bar{\theta}\geq 0)$} The only difference with respect to the previous sub-case  is that the expansion of the intersecting ingoing radial null geodesics does not remain negative, $\bar{\theta}=\left.\theta_-\right|_{\lambda \rightarrow \infty}\geq0$.
\end{itemize}

\item[Case C]{No defocusing point, $\lambda_\defocus=\emptyset$.}
\begin{itemize}
\item[C.I:]{$(\emptyset,0,\bar{\theta}<0)$: 
The expansion $\theta_+$ remains negative (but finite) in the limit $\lambda\rightarrow\infty$, which in particular implies that the integral in Eq. \eqref{First} is divergent. Thus, the radius of the area element orthogonal to the congruence vanishes asymptotically along these geodesics. The expansion of the intersecting ingoing radial null geodesics remains negative, so that $\bar{\theta}=\left.\theta_-\right|_{\lambda \rightarrow \infty}<0$.}

\item[C.II:]{$(\emptyset,0,\bar{\theta}\geq0)$: 
The only difference with respect to the previous sub-case  is that the expansion of the intersecting ingoing radial null geodesics does not remain negative, $\bar{\theta}=\left.\theta_-\right|_{\lambda \rightarrow \infty}\geq0$.}
\end{itemize}
\end{description}

We can now discuss these different cases separately {in order to understand in more detail the properties of these spacetimes and, in particular, the implications of the homogeneity and isotropy assumptions}. Given the quite different nature of {closed universes,} we shall treat first the flat and open cases in Sec.~\ref{sec:flat} and~\ref{Sec:open}, and consider separately the closed case in Sec.~\ref{sec:Closed}.
 
\section{Regular Flat FLRW universes} \label{sec:flat}

We shall now discuss in detail the different cases discussed in Sec.~\ref{taxonomy} for a flat FLRW universe. We will be using often the expression for the expansion of ingoing and outgoing null geodesics, Eq.~\eqref{expansion}, specialized to $k=0$ so that $R(t)=r(t)$:
\begin{equation}
 \theta_{\pm} =\frac{2}{a^2(t)} \left[ -\dot{a}(t) \pm \frac{1}{r(t)}\right]=\frac{2}{a^2(t) r(t)} \left[ -\dot{a}(t) r(t) \pm 1\right]\,.
 \label{eq:expansion_k0_1}
\end{equation}

In order to have a more clear physical interpretation of this expression, it is useful to rewrite it as
\begin{equation}
 \theta_{\pm} =\frac{2}{a(t)} \left[ -\frac{\dot{a}(t)}{a(t)} \pm \frac{1}{a(t) r(t)}\right]=\frac{2}{a(t)} \left[ -H(t)  \pm \frac{1}{a(t) r(t)} \right]\,.
 \label{eq:expansion_k0_2}
\end{equation}

We see that null rays for which the comoving radius $a(t) r(t)$ lies outside the Hubble sphere $1/H(t)$, are trapped. The Hubble sphere is the point in which the recession velocity due to the cosmological expansion equates the speed of light~\cite{Ellis:1993fkf,Davis:2003ad} (indeed, the Hubble radius is sometimes called the ``speed of light sphere"~\cite{Ellis:1993fkf}). Now, consider two past-directed outgoing {null} geodesics, thay have a tendency to move away from each other but, since they are past directed, these experience a contraction of the universe that tends to bring them closer. They will effectively move away from each other only if their velocity is larger than the ``contraction" velocity of the universe, that is, if their comoving radius is inside the Hubble sphere. The remaining null geodesics, lying outside $1/H(t)$, will inevitably get closer to each other driven by the ``contraction"  of the space-time, and are indeed trapped.

\subsection{Cases A.I and A.II $(\lambda_0,R_0$) }\label{SubSecCaseA}

For a given congruence of outgoing null geodesics that is trapped at $t=0$, it is required that $\theta_{+}<0$, which means that
\begin{equation}
    \dot{a}(0)>\frac{1}{r(0)} \rightarrow \dot{a}(0)r(0)>1.
\end{equation}
The existence of a defocusing point at a finite affine distance $t=t_0$ requires the condition
\begin{equation}\label{eq:defcond1}
\Dot{a}(t_0)=\frac{1}{r(t_0)} \rightarrow \Dot{a}(t_0)r(t_0)=1.
\end{equation}
Hence, a relative deceleration towards the past is required. Also, it is necessary that the decreasing scale factor does not vanish in the interval $t\in[0,t_0)$. Otherwise, we would have $\theta_+(t)=-\infty$, thus signaling the formation of a focusing point, which we were trying to avoid (the limiting case in which $a(t_0)\rightarrow0$ slow enough so that $\theta_+(t_0)\rightarrow0$ is discussed below). 
The condition in Eq.~\eqref{eq:defcond1} must be satisfied for some value of $t_0$ along all congruences. As shown next, this imposes a stronger constraint on $\dot{a}(t)$.

{\bf Proposition:}
All congruences of null geodesics are untrapped at finite affine distance $\lambda$ corresponding to time $t$ if and only if the derivative of the scale factor with respect to the time, $\Dot{a}(t)$, vanishes.

\textbf{Proof:} Let us assume that $\Dot{a}(t)$ does not vanish for any value of $t$, and show that this implies the existence of at least a congruence of past-directed null geodesics that does not have a defocusing point at finite affine distance. Let us take a reference congruence, for which
\begin{equation}
1-\Dot{a}(0)r(0)<0,
\end{equation}
thus implying that $\theta_+(0)<0$, while 
\begin{equation}
1-\Dot{a}(t_0)r(t_0)
\end{equation}
has an indefinite sign. Let us now multiply $r(t)$ by a positive constant factor $N$, which yields another congruence with a larger value of the radius. We still have that
\begin{equation}
1-\Dot{a}(0)Nr(0)=1-\Dot{a}(0)r(0)-(N-1)\Dot{a}(0)r(0)<0,
\end{equation}
so this second congruence was still trapped originally. Moreover, it always exists a value of $N$ such that
\begin{equation}
1-\Dot{a}(t_0)Nr(t_0)<0.
\end{equation}
This holds for any finite value of $t_0$, thus showing that there always exists a congruence of past-directed outgoing null geodesics that remains trapped, thus reaching a contradiction.

Regarding the converse implication, if we impose $\Dot{a}(t_0)=0$, then the expansion of any past-directed congruence of null geodesics is positive. $\qedsymbol$

Hence, spacetimes in this class are geodesically complete if and only if the derivative of the scale factor vanishes at some $t=t_0$, while the scale factor remains non-zero at $t=t_0$. Indeed, the limiting case in which also the scale factor vanishes is not admissible --- as it would entail a manifest curvature singularity. For $t<t_0$, in order to avoid the formation of further trapped surfaces, $\Dot{a}(t) \leq 0$. Hence, spacetimes in this class describe a bounce between a contracting and an expanding universe (\cref{Bounce}) where a future trapped region is continuously connected with a past trapped region, being both  delimited by the trapping horizon $r=\pm{1}/{a(t)}$. 
\begin{figure}[htb]
   \includegraphics[width=7cm]{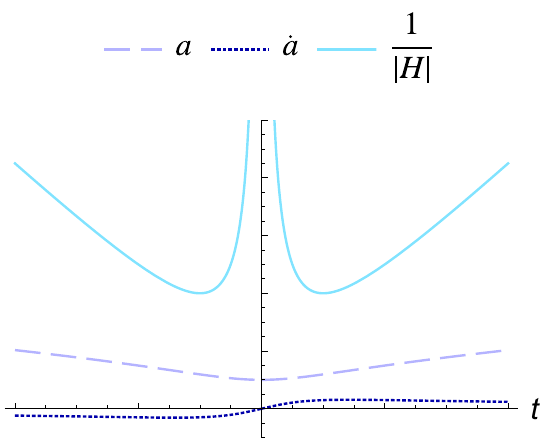}
    \caption{\small Bouncing universe: behaviour of the scale factor $a(t)$, of its derivative to respect to time and of the Hubble radius ${1}/{|H(t)|}={a(t)}/{\dot{a}(t)}$. The defocusing point $\dot{a}(t)=0$ is reached in a finite affine distance and it is preceded by a ``contracting" phase in which $\dot{a}(t)<0$.}
    \label{Bounce}
\end{figure}

Note that we are implicitly assuming that the scale factor is an analytic function of time which, if dropped, would allow also the possibility that $a(t)$ vanishes in a closed interval or half-line (in the latter case, the geometry describes an expanding universe emerging from a stationary phase~\cite{Ellis:2002we}). Analytical solutions that are effectively emergent are however possible, for example the expansion can be preceded by an oscillatory phase in which $\dot{a}$ is 0 on average (see \cite{Alesci:2016xqa} {for a specific realization}). These solutions embody a hybrid nature, drawing from both emergent and bouncing universe scenarios, as $\dot{a}$ assumes also negative values during the ``emergent" phase (\cref{Emerg}).

The causal structure of the spacetimes entering this class is shown in \cref{BouncePen}.
\begin{figure}[h]
\begin{minipage}[b]{15.5cm}
    \includegraphics[width=15cm]{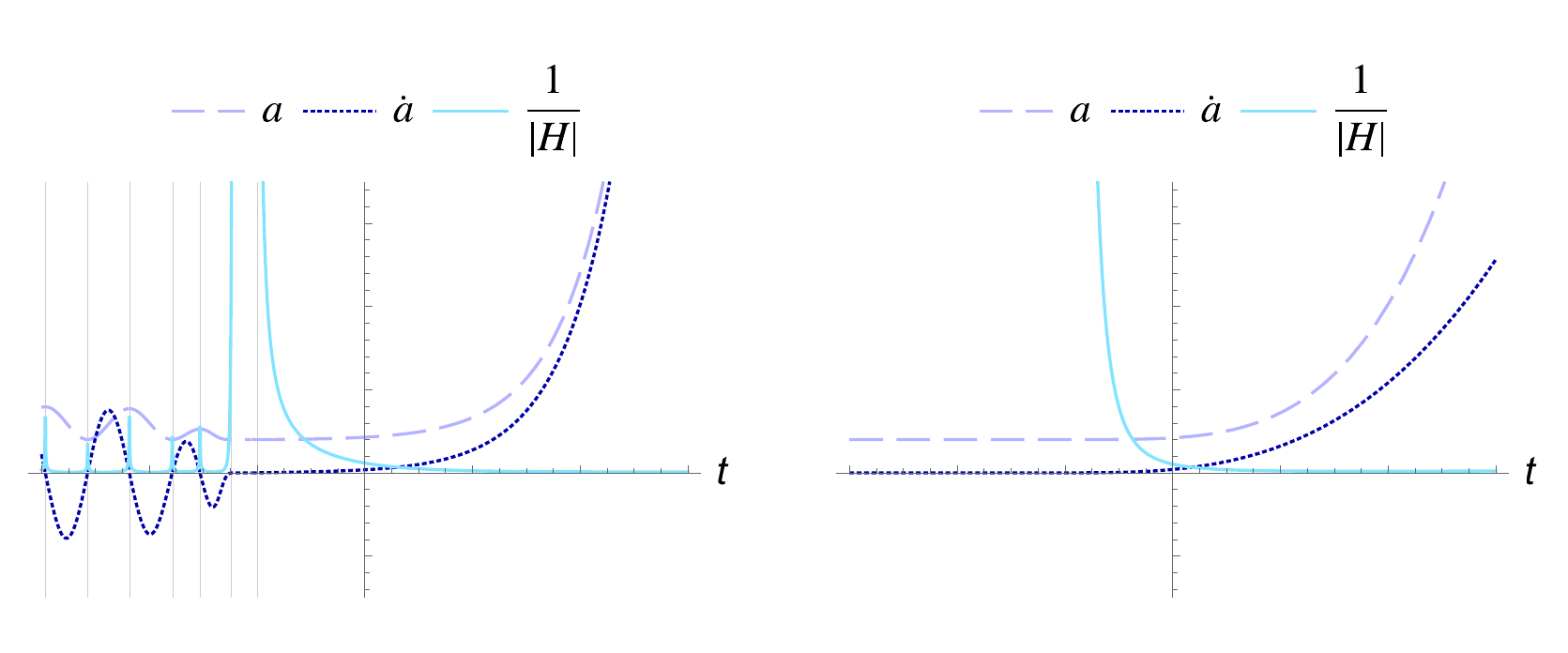}
    \caption{\small Effectively emergent universe and emergent universe: behaviour of the scale factor $a(t)$, of its derivative to respect to time and of the Hubble radius ${1}/{|H(t)|}={a(t)}/{\dot{a}(t)}$. The defocusing point $\dot{a}(t)=0$ is reached in a finite affine distance and it is preceded either by a ``stationary" phase in which $\dot{a}(t)$ is 0 on average (in the effectively emergent case --- left panel) or exactly 0 (in the emergent case --- right panel).}
    \label{Emerg}
    \ \hspace{200mm} \hspace{20mm}\
    \end{minipage}
   \end{figure}

As we {have mentioned above,} the expansion does not vanish at the same value of the affine parameter for all null outgoing geodesics. Furthermore, the homogeneity and isotropy assumptions strongly constrain the behaviour of ingoing geodesics with respect to the outgoing ones. Indeed, if $\theta_+(t)=0$, then
\begin{equation}
\theta_-(t)=-\frac{4}{a^2(t)r(t)}\leq0.    
\end{equation}

Hence, it is not possible to achieve at the same time $\theta_+(t) = 0$ and $\theta_-(t)\geq 0$ for all null geodesics (Case A.II) within the family of geometries being considered.

\begin{figure}
    \centering
\includegraphics[width=4cm]{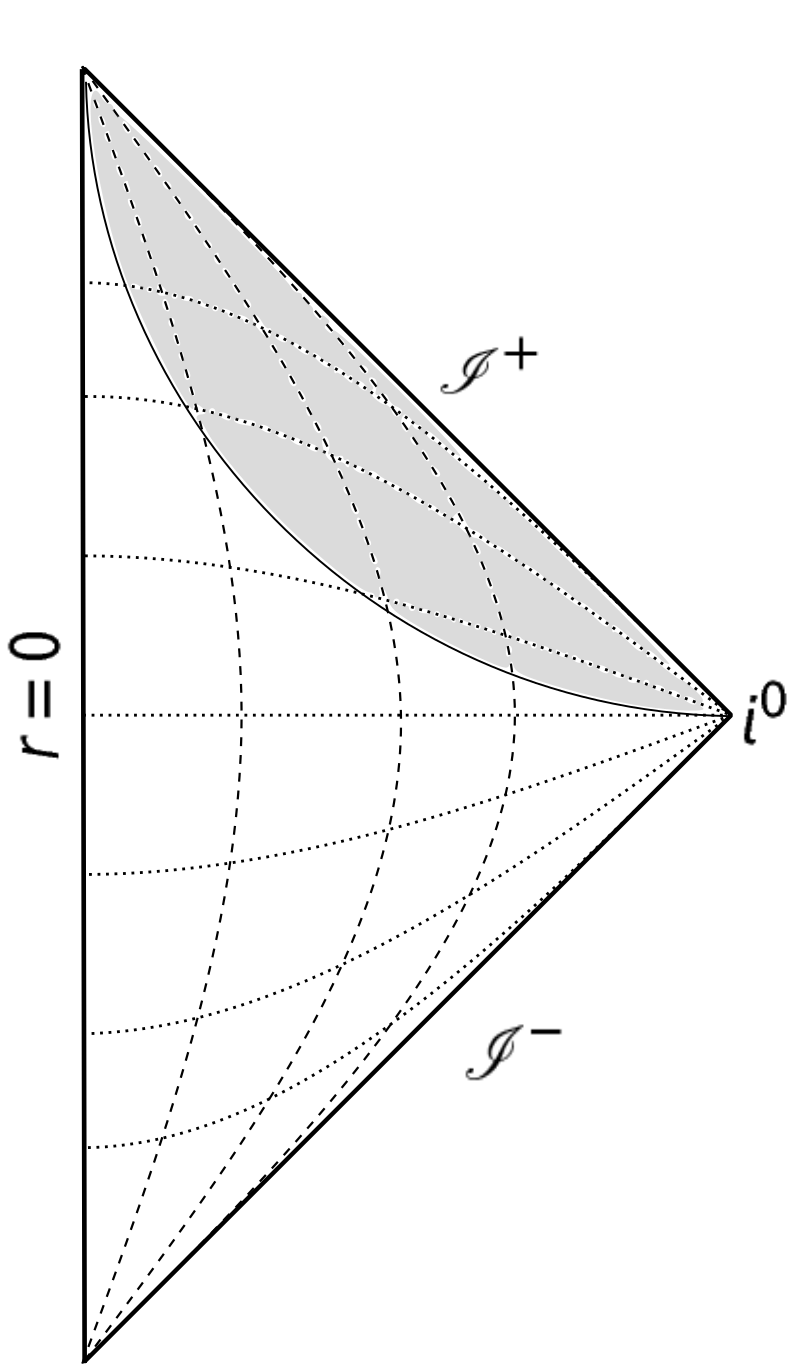}
    \caption{\small Penrose Diagram of a bouncing or emergent universe.}
    \label{BouncePen}
\end{figure}
An analogous reasoning leads to the same conclusions also for the case B.II, B.IV and C.II, thus we neglect this discussion about ingoing geodesics in the next sections.

\subsection{Cases B.I and B.II $(\infty,R_{\infty})$}
This case is similar to A.I, but now the defocusing point is at an infinite affine distance $\lambda_{\rm defocus}=\infty$. Since $a(t)$ is always finite, this means that $t_{\rm defocus}=-\infty$. We have that the integral
\begin{equation}
\lambda_{\rm defocus}=\lambda_*+2\int_{-\infty}^0\text{d}t\,a(t)
\end{equation}
is divergent, while the integral
\begin{equation}
\ln\left(\frac{\left.\delta A_+\right|_{\lambda=\infty}}{\left.\delta A_+\right|_{\lambda=\lambda_*}}\right)=\int_{-\infty}^{0}\text{d}t\,a(t)\theta_+(t)=\int_{-\infty}^{0}\text{d}t \frac{2}{a(t)}\bigg(-\dot{a}(t)+\frac{1}{r(0)+\int_{t}^{0}\text{d}t'/a(t')}\bigg)
\label{Int}
\end{equation}
must not be negatively divergent ($> -\infty$), where we have used the expression $r(t)=r(0)+\int_{t}^{0}\text{d}t'/a(t')$ obtained along geodesics, with tangent vector given by Eq.~\ref{geo}.

\textbf{Proposition:} All congruences of null geodesics are untrapped at infinite
affine distance   if and only if the derivative of the scale factor with
respect to the time,  $\dot{a}(t)$, vanishes for $t \rightarrow-\infty$.

\textbf{Proof:} The proof proceeds as the one for Case A.I simply putting $t_0=-\infty$. $\qedsymbol$

\textbf{Example:} Let us require to have a regular expanding universe at any time thus $\Dot{a}(t)\geq 0$ and $a(t)>0$ for any $t$. In order to have the defocusing at infinite affine distance we then need $\lim_{t \rightarrow -\infty}\Dot{a}(t)= 0$. {A simple analytic profile} for $a(t)$ is then $ a(t)=a_0+b_0 e^{H t}$, that is a sort of ``regularized inflation" (see \cref{Infl} and \cref{InflPen}). 
\begin{figure}[htb]
   \includegraphics[width=7cm]{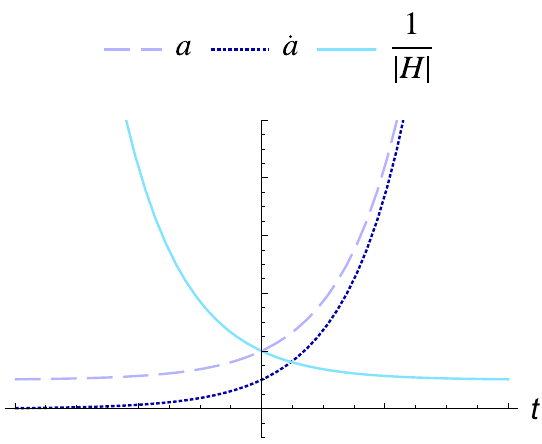}
    \caption{\small Regular inflating universe: behaviour of the scale factor $a(t)$, of its derivative to respect to time and of the Hubble radius ${1}/{|H(t)|}={a(t)}/{\dot{a}(t)}$. The defocusing point $\dot{a}(t)\rightarrow0$ is reached in an infinite affine distance. }
    \label{Infl}
\end{figure}
With this choice {the integral in Eq.~\eqref{Int} goes to $+ \infty$. It is possible to prove that this is a generic feature of each regular spacetime entering this case.}

\textbf{Proposition:} For a regular expanding flat FLRW spacetime with complete defocusing at infinite affine distance [$\dot{a}(t)$ vanishes for $t \rightarrow-\infty$] the integral in Eq.~\eqref{Int} diverges positively, $\ln\left({\left.\delta A_+\right|_{\lambda=\infty}}/{\left.\delta A_+\right|_{\lambda=\lambda_*}}\right)=+ \infty$.

\textbf{Proof:}
If the space-time is regular, $a(t)\neq 0$ $\forall t$, then the first term in the integral Eq.~\eqref{Int},
\begin{equation}
    \int_{-\infty}^{0}\text{d}t \bigg( -2 \frac{\dot{a}(t)}{a(t)} \bigg),
\end{equation}
is negative but finite. Indeed, if $a(t)\neq 0$ the integrand in the equation above is finite, thus the integral from $0$ to $t_B$ with $t_B$ arbitrarily small, but finite, cannot diverge. On the other hand,
Since $a(t)$ is a monotonically increasing function in the interval $t\in[- \infty,t_B]$ and it starts from a value $a(-\infty)>0$, we have
\begin{equation}
   \int_{-\infty}^{t_B}\text{d}t \bigg( -2 \frac{\dot{a}(t)}{a(t)} \bigg) >  \int_{-\infty}^{t_B}\text{d}t \bigg( -2 \frac{\dot{a}(t)}{a(-\infty)} \bigg)= -2 \bigg( \frac{a(t_B)}{a(-\infty)}-1\bigg) > - \infty.
   \label{finite}
\end{equation}

Let us therefore focus on the second term:
\begin{equation}
    \int_{-\infty}^{0} \frac{2\,\text{d}t}{a(t)\left[r(0)+\int_t^{0}{\text{d}t'}/{a(t')}\right]}.
\end{equation}
As mentioned before, we are choosing a reference point $\lambda_*$ corresponding to $t=0$ at which $\delta A$ is finite\footnote{This is always possible through a linear rescaling of the time coordinate $t \rightarrow t+c$ with $c$ an arbitrary constant.} and thus $a(t=0)=a_0$ is a finite positive number. 
Since $a(t)$ is a {monotonically} increasing function from $-\infty$, $a(t)<a_0$ in all the interval $[0,-\infty)$ thus
\begin{equation}
    \int_{-\infty}^{0}{ \frac{2\,\text{d}t}{a(t)\left[r(0)+\int_t^{0}{\text{d}t'}/{a(t')}\right]}}>\int_{-\infty}^{0}{ \frac{2\,\text{d}t}{ a_0\left[r(0)+\int_t^{0}{\text{d}t'}/{a(t')}\right]}}.
\end{equation}
We can split this integral into two pieces by defining a negative value of $t$, $t_{\rm B1}$,
\begin{equation}
   \int_{-\infty}^{0} \frac{2\,\text{d}t}{a_{0}\left[r(0)+\int_t^{0}{\text{d}t'}/{a(t')}\right]}=\int_{t_{\rm B1}}^{0} \frac{2\,\text{d}t}{a_{0}\left[r(0)+\int_t^{0}{\text{d}t'}/{a(t')}\right]}+  \int_{-\infty}^{t_{\rm B1}} \frac{2\,\text{d}t}{a_{0}\left[r(0)+\int_t^{0}{\text{d}t'}/{a(t')}\right]}, 
\end{equation}
with the first term being finite, so that any divergent behavior (if present) is isolated in the second term. Let us therefore focus on this second term, introducing another auxiliary value of the time $t_{\rm B2}>t_{\rm B1}$ and expanding $a(t)$ near $-\infty$ as $a(t)=a_{-\infty}+a_{1}/t + a_{2}/t^2+...$, so that
\begin{align}
&\int_{-\infty}^{t_{\rm B1}}\frac{2\text{d}t }{a_{0}\left[r(0)+\int_t^{0}{\text{d}t'}/{a(t')}\right]}\nonumber\\
   &=\int_{-\infty}^{t_{\rm B1}} \frac{2\text{d}t}{a_{0}\left[r(0)+\int_{t_{\rm B2}}^{0}{\text{d}t'}/{a(t')}+\int_t^{t_{\rm B2}}{\text{d}t'}/\left(a_{-\infty}+a_{1}/{t'} + a_{2}/{t'^2}+\dots\right)\right]}\nonumber\\
   &=\int_{-\infty}^{t_{\rm B1}} \frac{2\text{d}t}{a_{0}\left[r(t_{\rm B2})+\int_t^{t_{\rm B2}} \text{d}t' \left({1}/{a_{-\infty}}+b_1/{t'}+b_ 2/{t'^2}+\dots\right) \right]}\nonumber\\
   &=\int_{-\infty}^{t_{\rm B1}} \frac{2\text{d}t}{a_{0}\left[g(t_{\rm B2})- {t}/{a_{-\infty}} - b_1 \ln{|t|}+ b_2/{t}+\dots \right]},
\end{align}
where we introduced $g(t_{\rm B2})=r(t_{\rm B2})+ {t_{\rm B2}}/{a_{-\infty}} + b_1 \ln{|t_{\rm B2}|}- b_2/{t_{\rm B2}}$ that is not a function of $t$. Since the integrand goes to 0 in the $t\rightarrow-\infty$ limit at most as ${1}/{t}$, the integral in the last line is divergent. 

\begin{figure}
    \centering
\includegraphics[width=3.5cm]{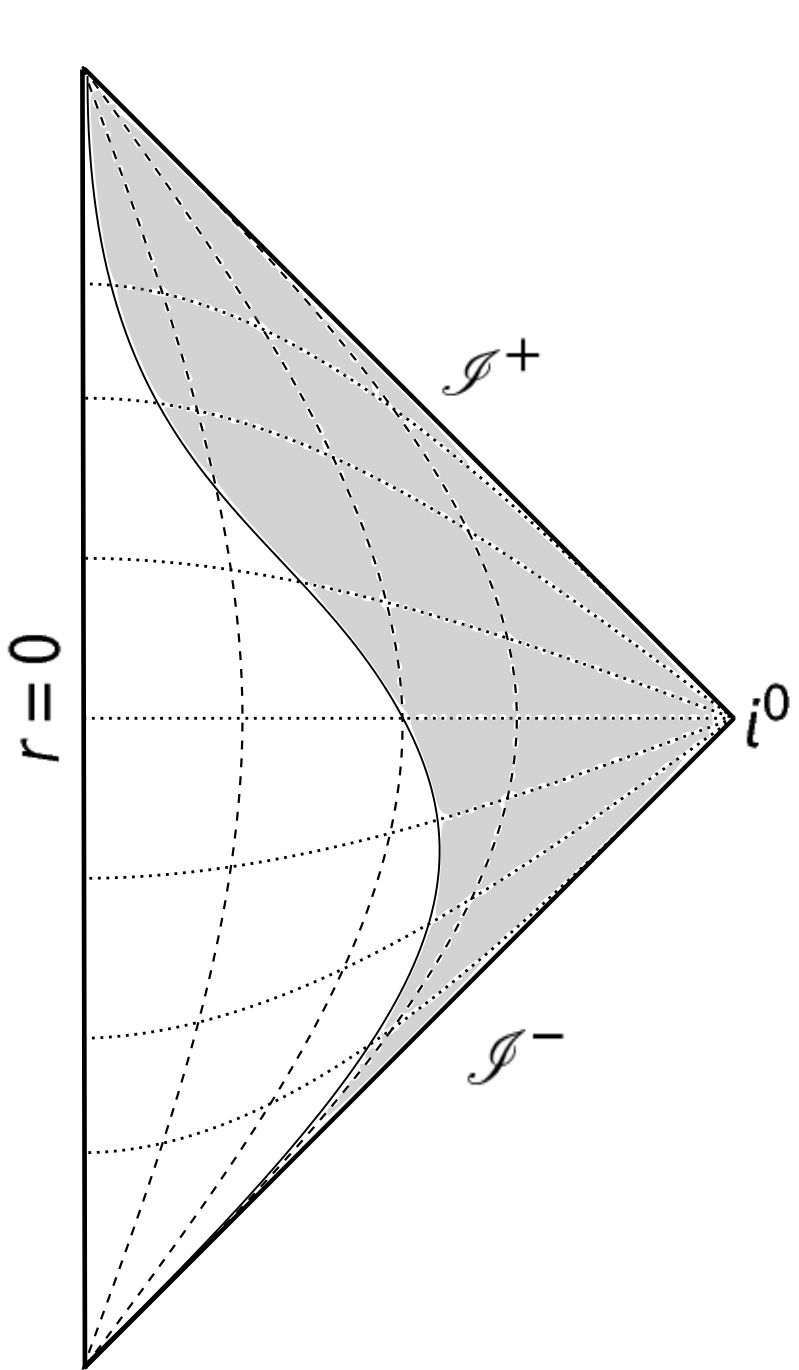}
    \caption{\small Penrose Diagram of a regular universe of Case B.I
    (asymptotic defocusing e.g. $ a(t)=a_0+b_0 e^{H t}$)}
    \label{InflPen}
\end{figure}

This analysis is general enough for being in principle applicable also to the cases B.I and B.II. Nonetheless, as explained at the end of Sec.~\ref{SubSecCaseA}, case B.II, cannot be realized, within the family of geometries being here considered, given that the conditions $\theta_+=0$ and $\theta_-\geq 0$ are never realized simultaneously for all null geodesics.

\subsection{Cases B.III and B.IV $(\infty,0)$}
In this case, the defocusing point is at an infinite affine distance and the integral in Eq.~\eqref{Int} is divergent ($-\infty$). As discussed below, this case is singular. 
 
{\bf Proposition:} If the integral in Eq.~\eqref{Int} is {negatively} divergent (it goes to $-\infty$), then $a(t) \rightarrow 0$ for some $t$, and thus the spacetime is also singular.
 
{\bf Proof:} 
We will prove the equivalent proposition: if $a(t)\neq 0$ $\forall t$ (including the limit  $t \rightarrow -\infty$), then the integral Eq.~\eqref{Int} {cannot be negatively divergent.} 
Indeed, the term proportional to $1/r(t)$ is always positive thus, to prove that the integral is greater than $-\infty$, we only need to focus on the following piece
\begin{equation}
   \int_{-\infty}^{0}\text{d}t \bigg( -2 \frac{\dot{a}(t)}{a(t)} \bigg).
\end{equation}
which however, as already proved, is never negatively divergent [see, in particular, Eq.~\eqref{finite} in the previous section].

Finally note, that the above reasoning is strictly needed for case B.III only, given that case B.IV, as explained {at the end of Sec.~\ref{SubSecCaseA}}, is not realizable within the family of geometries being considered.

\subsection{Cases C.I and C.II $(\emptyset,0)$}
In this case, there is some congruence for which  $\theta_+$ remains negative for $\lambda \rightarrow + \infty$ ($t \rightarrow -\infty$). This can happen only if $\dot{a}(t)$ remains positive in the infinite domain $(-\infty,0]$. Thus, if $a(t)$ is finite at the point $t=0$, it must have crossed or reached the value of 0 at some point in the past. This proves that there is a curvature singularity, and therefore these spacetimes cannot be regular.

Note that, as explained at the end of Sec.~\ref{SubSecCaseA}, aside from this singular behavior, it would not be possible to realize Case C.II within the family of geometries being considered.
\begin{figure}[htb]
    \centering
  \includegraphics[width=15 cm]{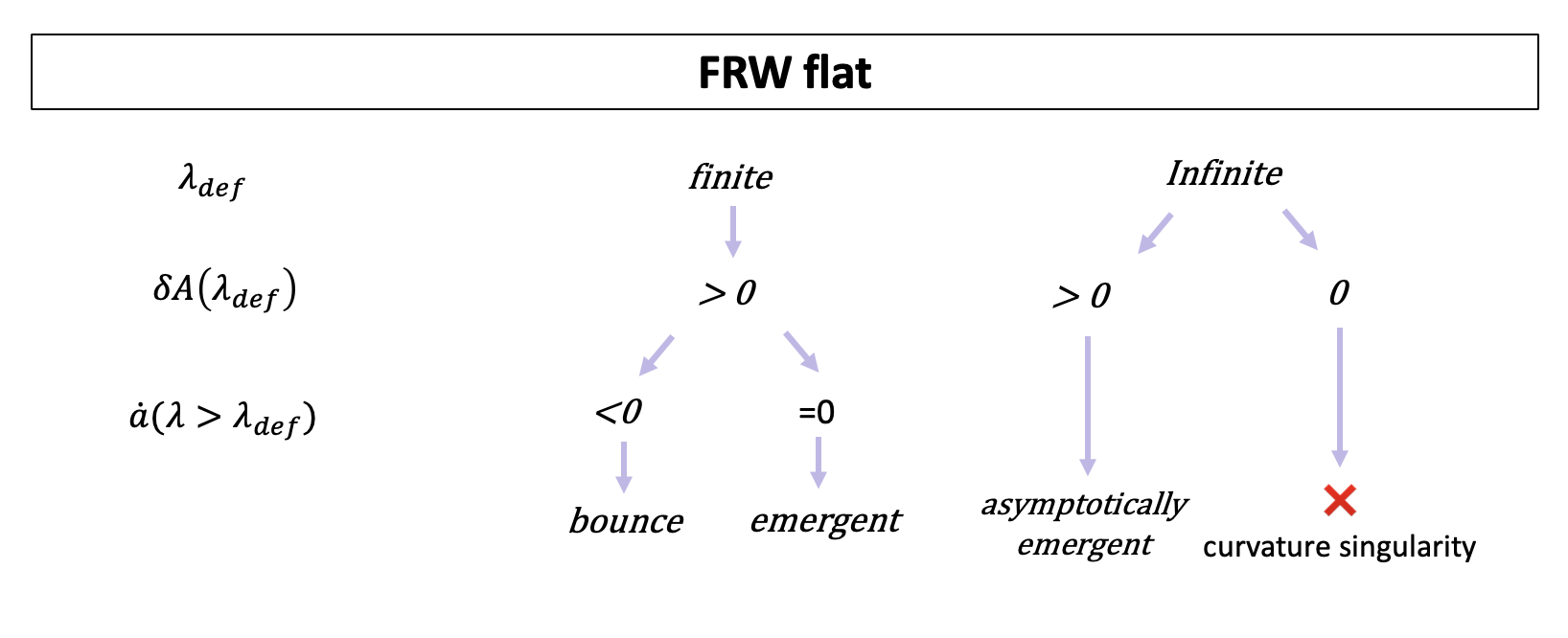}
    \caption{\small Scheme of the allowed flat homogeneous and isotropic regular geometries. These geometries are classified according to: i) the value of the affine parameter at which all outgoing null geodesics are defocused $\lambda_{def}$ so that $\theta_{+}(\lambda_{def})=0$; ii) the area element orthogonal to the congruence at $\lambda_{def}$; iii) the derivative of the scale factor for $\lambda>\lambda_{def}$ that is in the past with respect to the defocusing ($t<t_{def}$). }
    \label{SchemeFlat}
\end{figure}

\section{Regular Open FLRW universes}
\label{Sec:open}

We shall now discuss in detail the different cases discussed in Sec.~\ref{taxonomy} for a open FLRW universe. The expansion of ingoing and outgoing null geodesics, Eq.~\eqref{expansion}, particularized to $k=-1$, i.e.~$R(r)=\tan{r(t)}$, is
\begin{eqnarray}
 \theta_{\pm}(t) =\frac{2}{a^2(t)} \left[ -\dot{a}(t) \pm \frac{1}{\tanh{r(t)}}\right]=\frac{2}{a^2(t)\tanh{r(t)}} \left[1 -\dot{a}(t)\tanh{r(t)}\right].
 \label{eq:expansion_k0_3}
\end{eqnarray}

\subsection{Cases A.I and A.II $(\lambda_0,R_0)$}
\label{par:IngoOpen}
{\bf Proposition:}
All congruences of null geodesics are untrapped at finite affine distance $\lambda$ corresponding to time $t$ if and only if $\Dot{a}(t) \leq 1$.

\textbf{Proof:} 
Let us assume that $\Dot{a}(t)$ remains greater than $1$ for any value of $t$, and show that this implies the existence of at least a congruence of past-directed null geodesics that does not have a defocusing point at affine distance. Let us take a reference congruence, for which
\begin{equation}
1-\Dot{a}(0)\tanh{r(0)}<0,
\end{equation}
thus implying that $\theta_+(0)<0$, while 
\begin{equation}
1-\Dot{a}(t_0)\tanh{r(t_0)}=0
\end{equation}

Let us now multiply $r(t)$ by a positive constant factor $N$, which yields another congruence with a larger value of the radius. Then we have that:
\begin{equation}
    \tanh{Nr(t)}=M \tanh{r(t)}
\end{equation}
with $M>1$ since $\tanh(x)$ is  monotonous increasing in $[0,+ \infty)$.
Thus we still have that
\begin{eqnarray}
1-\Dot{a}(0)\tanh{Nr(0)}=1-2M \Dot{a}(0)\tanh{r(0)} 
= \nonumber\\
1-\Dot{a}(0)\tanh{r(0)}-\Dot{a}(0)(M-1)\tanh{r(0)}<0,
\end{eqnarray}
so this second congruence was still trapped originally. Moreover, it always exists a value of $N$ such that
\begin{equation}
1-\Dot{a}(t_0)\tanh{Nr(t_0)}=1- M \Dot{a}(t_0)\tanh{r(t_0)}<0.
\label{condition}
\end{equation}
This holds for any finite value of $t_0$, thus showing that there always exists a congruence of past-directed outgoing null geodesics that remains trapped, thus reaching a contradiction.
Regarding the converse implication, if we impose $\Dot{a}(t_0) \leq 1$, then $\Dot{a}(t_0) \tanh{ r(t_0)} \leq \tanh{r(t_0)} \leq 1$ 
and thus the expansion of any past-directed congruence of null geodesics is positive. $\qedsymbol$

{Note that it is possible to have singular spacetimes without trapped regions if $a(t)$ vanishes, for instance $a(t)=e^t$. In this case there is a naked curvature singularity at $t=-\infty$ where $\theta_{\pm} = \pm \infty$, which is therefore not a focusing point. Hence, the condition $\dot{a}\leq1$ is necessary, but not sufficient for regularity.}

As in the flat case, the homogeneity and isotropy assumptions strongly constrain the behaviour of ingoing geodesics with respect to the outgoing ones. Indeed, if $\theta_+(t)=0$, then
\begin{equation}
\theta_-(t)=-\frac{4}{a^2(t) \tanh{r(t)}}\leq0.   
\end{equation}
Hence, it is not possible to achieve at the same time $\theta_+(t) = 0$ and $\theta_-(t)\geq 0$ for all null geodesics within the family of geometries being considered, and therefore case A.II cannot be realized.

An analogous reasoning leads to the same conclusions for cases B.II, B.IV and C.II, thus we will not repeat explicitly this discussion about ingoing geodesics in the next sections.

\subsection{Cases B.I and B.II $(\infty,R_{\infty})$}
In this case, the defocusing point is at an infinite affine distance, and the integral in Eq.~\eqref{Int} is convergent.

{\bf Proposition:}
This case cannot be realized, if you push the defocusing point to $\infty$ than the integral in Eq.~\eqref{Int} diverges.

{\bf Proof:}
To have the defocusing point at $\infty$, $\dot{a} \rightarrow 1$ for $t \rightarrow -\infty$. This can be seen following the proof of Case A.I simply putting $t_0=-\infty$. 
If $\dot{a} \rightarrow 1$ for $t \rightarrow -\infty$ it always remains positive in the infinite domain $(-\infty, t(0))$, thus  if $a(t)$ is finite at the point $t=0$, it must have crossed or reached the value of 0 at some point in the past. This proves that the integral in Eq.~\eqref{Int} diverges and that there must be a curvature singularity.

As explained {at the end of Sec.~}\ref{par:IngoOpen}, Case B.II can not be achieved within the family of geometries being considered.

\subsection{Cases B.III and B.IV $(\infty,0)$}
The defocusing point is at an infinite affine distance and the integral in Eq.~\eqref{Int} is divergent (it goes to $- \infty$). 

{\bf Proposition:} This case is singular.

{\bf Proof:} 
If $\dot{a} \rightarrow 1$ for $t \rightarrow -\infty$ it always remain positive in the infinite domain $(-\infty, t(0))$, thus  if $a(t)$ is finite at the point $t=0$, it must have crossed or reached the value of 0 at some point in the past. This proves that there must be a curvature singularity. 

Note that, as explained {at the end of Sec.~}\ref{par:IngoOpen}, it is not possible to achieve Case B.IV within the family of geometries being considered.

\subsection{Cases C.I and C.II $(\emptyset,0)$}
This case is singular, and the proof proceeds as for case B.III.

As discussed {at the end of Sec.~}\ref{par:IngoOpen}, it is impossible to achieve case C.II within the family of geometries being considered.

\begin{figure}[htb]
    \centering
  \includegraphics[width=15cm]{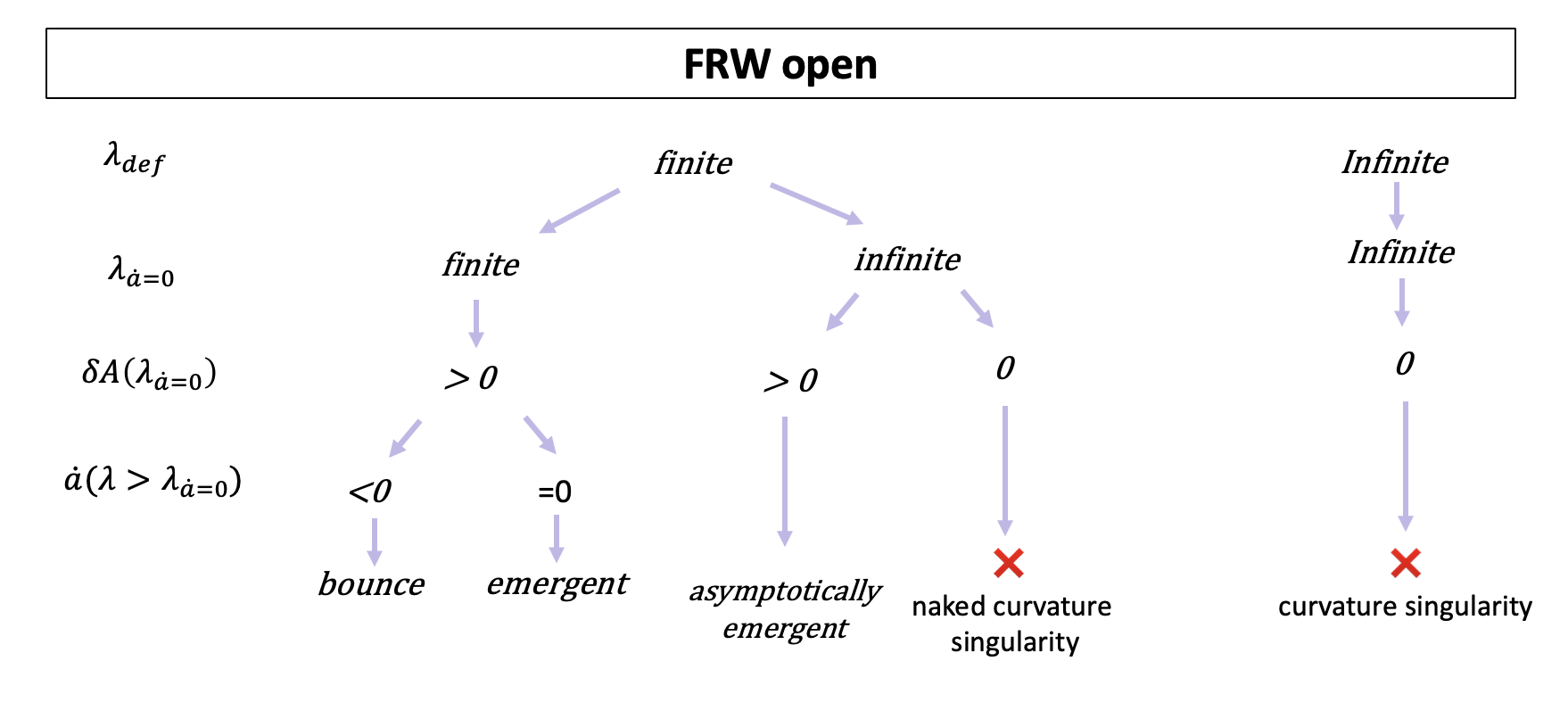}
    \caption{\small Scheme of the allowed open homogeneous and isotropic regular geometries. These geometries are classified according to: i) the value of the affine parameter at which all outgoing geodesics are defocused $\lambda_{def} | $ $\theta_{+}(\lambda_{def})=0$; ii) the value of the affine parameter at which the universe cease to contract/expand that is $\lambda_{\dot{a}=0}$; iii) the area element orthogonal to the congruence at $\lambda_{def}$; iv) the derivative of the scale factor for $\lambda>\lambda_{def}$, i.e.~for $t<t_{def}$. }
    \label{SchemeOpen}
\end{figure}

\section{Regular Closed FLRW universes}
\label{sec:Closed}

In the case of a closed universe, the non-compactness assumption of the Penrose theorem is no more guaranteed, thus in general we would have to rely on the Hawking-Penrose theorem. However, as we will show in the following, in the specific case of an homogeneous and isotropic universe (describable by the FLRW metric), simpler assumptions suffice to prove the presence of a singular focusing point. 

The first part of Penrose's theorem makes no use of the presence of a non-compact Cauchy hypersurface, and thus is also valid for any generic closed space-time, thus showing that an initial negative expansion and the null convergence condition are enough to prove the presence of a focusing point in which the expansion will become infinitely negative. What is no more guaranteed for closed space-times is that this focusing point is singular.

The expansion of null geodesics in a closed FLRW space-time is given by:
\begin{equation}\label{eq:exp_clo}
 \theta_{\pm} =\frac{2}{a^2(t)} \left[ -\dot{a}(t) \pm \frac{1}{\tan{r(t)}}\right].
\end{equation}
Note that $1/\tan{r}$ is a periodic function thus we can evaluate it in $r \in [0,\pi] $, taking values $\infty$ for $r \rightarrow 0$, $0$ for $r \rightarrow \pi/2$ and $-\infty$ for $r \rightarrow \pi$. Both poles (0 and $\pi$) correspond to $R(r)=0$, and the reason for which the expansions there blow up with different signs is geometrical and essentially the same reason for which the expansion of geodesics in flat Minkowski blows up at $r=0$. These points are then regular focusing points.

Using this expression for the expansions we now show that, in presence of a trapped region where both expansions $\theta_{\pm}$ are negative, a singular focusing point is always present if the timelike convergence condition holds (this implies the strong energy condition  in GR). In order to isolate the divergent behavior at the focusing point, while avoiding the divergences caused by the poles, it is useful to introduce the following quantity:
\begin{equation}
 \theta = \theta_{+}+\theta_{-}=-\frac{2 \dot{a}(t)}{a^2(t)}.
 \label{theta}
\end{equation}
If both $\theta_{\pm}$ are negative initially, and thus a trapped surface is present, then also $\theta$ will have an initial negative value $\theta_0<0$.
On the other hand, $\theta$ is the expansion of a past-directed time-like congruence of geodesics with tangent vector $U^{a}=(-2,0,0,0)/a(t)$ and, as such, it satisfies the Raychaudhuri equation with zero twist and shear:
\begin{equation}
   \frac{d \theta}{d \lambda}=-\frac{\theta^2}{3}-R_{ab}U^aU^b. 
\end{equation}
From the timelike convergence condition ($R_{\mu\nu}u^\mu u^\nu \geq 0 \quad \forall u^\mu$ timelike) --- or the strong energy condition if one assumes the Einstein field equations --- we have that
\begin{equation}
     \frac{d \theta_{\pm}}{d \lambda} \leq -\frac{\theta_{\pm}^2}{3},
\end{equation}
so that, in the best case scenario (the expansion taking the least negative value possible), we have
\begin{equation}
     \frac{d \theta}{d \lambda} = -\frac{\theta^2}{3}.
\end{equation}
Solving this differential equation results in the expression
\begin{equation}
    \theta^{-1}=\theta_0^{-1}+\frac{\lambda}{3},
\end{equation}
which indicates that $\theta$ will reach $-\infty$ for a finite value of the affine parameter (in particular, $-3\theta_0^{-1}$). 

On the other hand, taking a look at the functional form of $\theta$ in Eq.~\eqref{theta}, we see that it can reach $-\infty$ only if $a(t)=0$ or $\dot{a}(t)\rightarrow+\infty$. Both cases correspond to a divergence of the Ricci scalar in Eq.~\eqref{eq:Ricciscalar} and thus this focusing point is associated with a curvature singularity.

In summary, we have proved that a space-time satisfying the following assumptions cannot be geodesically complete:
\begin{itemize}
    \item It is an homogeneous and isotropic solution of Einstein equations.
    \item The timelike convergence condition holds.
    \item At some point a closed past-trapped surface forms where both expansions have a maximum negative value.
\end{itemize}
Based on this result, let us continue with our aim of characterizing the different possible regular solutions with a trapped region, following the classification discussed in Section \ref{taxonomy}.

\subsection{Case A\label{SubSecCaseAClosed}}
The reasoning is similar to the correspondig case for flat spacetime but taking into account that now $\theta_+<0$ doesn't correspond necessarily to a trapped region, since outgoing and ingoing geodesics exchange roles for some values of $r$ (see Fig.~\ref{closed}). We thus have to check the sign of both $\theta_+$ and $\theta_-$, and a trapped region is indeed present if both are negative. 

\begin{figure}
    \centering
  \includegraphics[width=6cm]{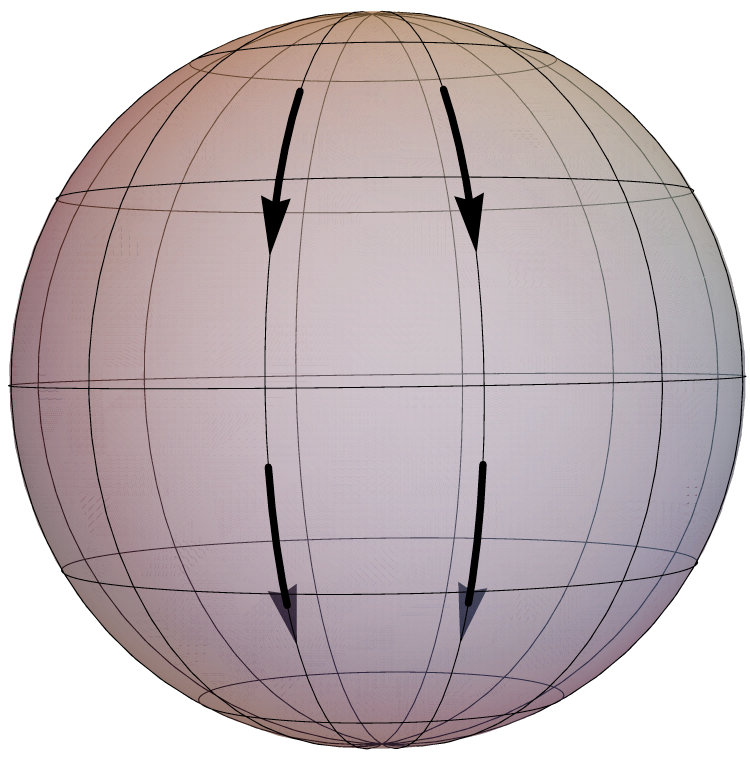}
    \caption{\small Path of geodesics in a 2-dimensional closed space-time. As we can see the in-going/outgoing nature of trajectories changes at the equator and the poles. }
    \label{closed}
\end{figure}

{\bf Proposition:}
All congruences of null geodesics are untrapped at finite affine distance $\lambda$ corresponding to time $t$ {\em if and only if} the derivative of the scale factor with respect to the time, $\Dot{a}(t)$, vanishes.

\textbf{Proof:}
 Let us assume that $\Dot{a}(t)$ does not vanish but remains positive for any value of $t$. Then for any $t$ all the geodesics for which $-\dot{a}(t)<1/\tan{r(t)}<\dot{a}(t)$ would have $\theta_\pm<0$ and thus would be trapped. 
These trapped geodesics always exist. Indeed since, for the periodic nature of the metric functions, $r$ can be taken in the finite interval $[0,\pi]$, it is always possible to chose an initial value for the photon position $r(0)$ such that $r(t)$ is sufficiently near to $\pi/2$ to satisfy the previous trapping condition. 
 Conversely, if $\dot{a}(t)$ becomes negative, $\theta_+<0$ only for geodesics with $1/\tan{r(t)}<\dot{a}(t)<0$ while $\theta_-<0$ only for geodesics with $1/\tan{r(t)}>-\dot{a}(t)>0$. Thus there are no geodesics for which both $\theta_+$ and $\theta_-$ are negative, and therefore no trapping region if $\dot{a}(t)$ becomes negative.

Hence, also for the closed case, this class describes a bounce with $\dot{a}(t)$ vanishing between a contracting and an expanding universe (or, if we drop the assumption that the scale factor is an analytic function of time, an expanding universe emerging from a stationary phase).

For what regards the distinction between case A.I and A.II, we saw that in a closed universe outgoing and ingoing geodesics exchange roles for some values of $r$ (see Fig.~\ref{closed}). 

\label{par:IngoClosed}
The homogeneity and isotropy assumption strongly constrain the behaviour of one family of geodesics with respect to the other. Indeed, if $\theta_+(t)=0$, then
\begin{equation}
\theta_-(t)=-\frac{4}{a^2(t)\tan r(t)}.    
\end{equation}
and if $\theta_-(t)=0$, then
\begin{equation}
\theta_+(t)=-\frac{4}{a^2(t)\tan r(t)}.    
\end{equation}
Hence, it is not possible to achieve at the same time $\theta_+(t) = 0$ and $\theta_-(t) \geq 0$ (or $\theta_-(t) = 0$ and $\theta_+(t) \geq 0$) for all null geodesics (case A.II) within the family of geometries being considered.

An analogous reasoning leads to the same conclusions also for the case B.II, B.IV and C.II thus we neglect this discussion about ingoing geodesics in the next sections.

\subsection{Case B} 
Analogously to the Case A, all congruences of null geodesics are untrapped at infinite
affine distance if and only if the derivative of the scale factor with
respect to the time,  $\dot{a}(t)$, vanishes for $t \rightarrow-\infty$. The proof proceeds as in Case A, simply considering $t\rightarrow-\infty$.

Differently from the flat and open case, we can no longer use the integral in Eq.~\eqref{First} to distinguish between cases B.I (no focusing point) and B.III (asymptotic focusing point) since in this case it always diverges for long enough geodesics (even for regular spacetimes), due to the divergent behaviour of the term $1/\tan{[r(t)]}$ in the expansion.
However, the presence of a singularity can be still detected in the behaviour of geodesics since it is signaled by the vanishing of the universe radius $a(t)$ that causes the negative divergence of both expansions, $\theta_{\pm} \rightarrow- \infty$. 

Note that, as explained in \ref{SubSecCaseAClosed}, it is not possible to achieve Case B.II and B.IV within the family of geometries being considered.

\subsection{Case C}
}

We can show that curvature invariants must be singular in this case. By definition, there is now some congruence for which both $\theta_\pm$ remain negative for $\lambda \rightarrow + \infty$ ($t \rightarrow -\infty$). {From Eq.~\eqref{eq:exp_clo}, we can conclude that} this can happen only if $\dot{a}(t)$ remains positive in the {semi-}infinite domain $(-\infty,0]$. {As a consequence,} if $a(t)$ is finite at $t=0$, it must have crossed or reached the value of $0$ at some point in the past {due to its derivative being definite positive in an semi-infinite domain}. This proves that there is a curvature singularity.

Note that, as explained in \ref{par:IngoClosed}, it is not possible to achieve Case C.II within the family of geometries being considered.

\begin{figure}[h]
    \centering
  \includegraphics[width=15cm]{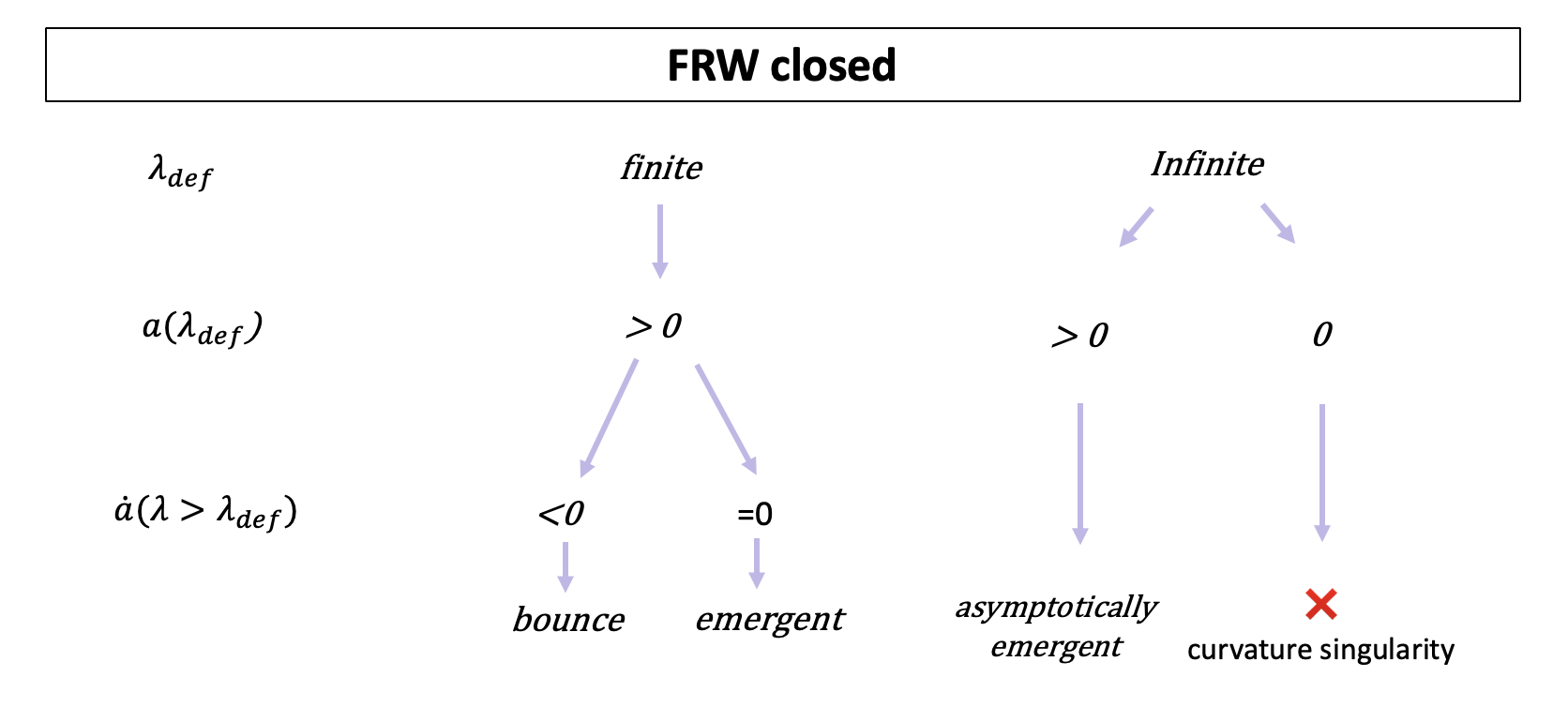}
    \caption{\small Scheme of the allowed closed homogeneous and isotropic regular geometries. These geometries are classified according to: (i) the value of the affine parameter at which all outgoing geodesics are defocused $\lambda_{def}$, so that $\theta_{+}(\lambda_{def})=0$; (ii) $a(\lambda_{def})$ or equivalently the value of both expansions for the same value of the affine parameter [note that, if $a(\lambda_{def})=0$, then $\theta_{\pm}(\lambda_{def})=-\infty$]; (iii) the derivative of the scale factor for $\lambda>\lambda_{def}$, i.e. for $t<t_{def}$.}
    \label{SchemeClosed}
\end{figure}

\section{Conclusions}
\label{sec:Concl}

This study delves into the behavior of null geodesics within various spacetime geometries in the context of cosmology. The investigation primarily focused on the behavior of the expansion of null congruences with the aim of classifying from a geometric standpoint the possible regularizations of the initial singularity. 
Distinct scenarios emerged based on the behavior of the expansion and the presence and extension of trapped regions. These scenarios are: \begin{itemize}
    \item A bouncing universe.
    \item An expanding universe emerging from a stationary phase.
    \item An asymptotically emergent universe where the scale factor is always decreasing towards the past but never vanishes. An example of scale factor with these characteristics is an inflating universe, with a characteristic exponential behavior of the scale factor but with the addition of a constant that corresponds to the asymptotic value of $a(t)$ towards the past.
\end{itemize} 
Some concrete examples for these three scenarios are constructed and studied in \cite{Easson:2024fzn} where it is explicitly shown that they all violate the null energy conditions for at least some amount of time.\\
The analysis of closed universes presented unique challenges due to the violation of the non-compactness assumption and to the diverging nature of certain functions influencing geodesic expansions. Despite this, it was established that, also in this case, true defocusing points at finite or infinite affine distance are contingent upon specific behaviors of the scale factor derivative with respect to time. Moreover, the absence of defocusing leads to the inevitable occurrence of a curvature singularity.
Regarding the physical interpretation of the bouncing solutions, {it is interesting to keep in mind the following consideration}. The metric alone does not provide any real information regarding the direction in which time flows. GR is time-symmetric, as flipping the $t$-direction on a globally hyperbolic manifold results in another valid solution to the field equations.\footnote{Note that, due to the fact that at the bounce $\dot{a}(t)=0$, this would be possible while keeping the metric components at least $C^1$, and could be possibly fine-tuned so that at the bounce also has $\Ddot{a}(t)=0$ so as to ensure regularity of curvature tensors as well.}

The scenario of a universe contracting to a minimum scale factor and then expanding stems from the implicit assumption that the direction of time remains the same before and after the bounce. However, it is possible to imagine a scenario in which the two parts of spacetime separated by the defocusing point have opposite time directions. In this case, we would have two identical expanding bouncing universes joined at the defocusing point (the bounce). The reason why we expect the inversion of time direction to be possible at the bounce is because, besides being a stationary point for the expansion of the universe, it is also a point at which we expect quantum gravitational effects to play an important role, causing this possible inversion. If for example, as in the Hartle-Hawking no boundary proposal \cite{Hartle:1983ai}, near the would-be-singularity time becomes imaginary, it would lose any privileged direction there, making the subsequent appearance of an opposite arrow of time more natural. Furthermore, this scenario seems also to connect to similar ideas recently advanced to address the so called ``arrow of time" problem~\cite{Barbour:2014bga}.

In conclusion, this study elucidates the diverse range of regular possibilities within cosmological models, shedding light on the interconnections between geodesic behavior, singularities, and the evolution of the universe.
These findings not only contribute to our theoretical understanding of the universe's behavior but also pave the way for refining cosmological models based on observational data and theoretical considerations.
Efforts to reconcile these theoretical studies with empirical observations and astrophysical data would be instrumental in refining and validating cosmological models.

\acknowledgments
{R. Carballo-Rubio acknowledges support from VILLUM Fonden through grant no.~29405.}
The authors would also like to thank Diego Buccio and Daniele Oriti for useful discussions.

\bibliographystyle{apsrev4-1}
\bibliography{main}
\end{document}